\documentclass[8pt]{article}  \usepackage{times}
\usepackage{graphicx}

\usepackage{color}

\usepackage[round,numbers,sort&compress]{natbib} 

\usepackage{amsmath}

\begin{document}

\twocolumn[{\LARGE \textbf{Voltage-gated lipid ion channels\\*[0.2cm]}}
{\large Andreas Blicher and Thomas Heimburg$^{\ast}$\\*[0.1cm]
{\small $^1$Niels Bohr Institute, University of Copenhagen, Blegdamsvej 17, 2100 Copenhagen \O, Denmark}\\

{\normalsize ABSTRACT\hspace{0.5cm} Synthetic lipid membranes can display channel-like ion conduction events even in the absence of proteins. We show here that these events are voltage-gated with a quadratic voltage dependence as expected from electrostatic theory of capacitors. To this end, we recorded channel traces and current histograms in patch-experiments on lipid membranes. We derived a theoretical current-voltage relationship for pores in lipid membranes that describes the experimental data very well when assuming an asymmetric membrane. We determined the equilibrium constant between closed and open state and the open probability as a function of voltage. The voltage-dependence of the lipid pores is found comparable to that of protein channels. Lifetime distributions of open and closed events indicate that the channel open distribution does not follow exponential statistics but rather power law behavior for long open times. \\*[0.0cm] }}
]

\noindent\footnotesize {$^{\ast}$corresponding author, theimbu@nbi.dk}\\


\normalsize
\section*{Introduction}
Synthetic lipid bilayers can display channel-like conduction \linebreak events through pores in the bilayer \cite{Heimburg2010}. This finding was first reported by Yafuso and collaborators \cite{Yafuso1974} on oxidized cholesterol membranes. Numerous subsequent reports show that one finds channel-like events in synthetic lipid membranes in the complete absence of proteins \cite{Antonov1980, Kaufmann1983a, Kaufmann1983b, Gogelein1984, Antonov1985, Kaufmann1989e, Antonov1990, Antonov2005}. The channels reported in those studies are typically of similar conductance (several 10 to 100 pS) and lifetime (a few ms to several 100 ms) than those of protein channels.  In various studies in the recent decade, Colombini and collaborators  reported that ceramide lipids may form pores in membranes with channel-like conduction appearance \cite{Siskind2000, Siskind2002, Siskind2003, Siskind2006a, Colombini2010, Perera2012}. The authors propose that ceramide channels are stable structures (comparable to a porin, see tentative structure in \cite{Shao2012}). The conductances of such pores are much larger (several nS), the lifetimes are longer (several seconds to minutes) and it is therefore uncertain whether ceramide-channels possess the same characteristics as other lipid pores, which rather resemble fluctuations in membrane density with amplitudes and lifetimes given by the thermal fluctuations of the lipid matrix (see below).

In the recent years, our own group has focused on the thermodynamics of membranes and in particular on the aspects of thermal fluctuations. Close to melting transitions the heat capacity of membranes is proportional to the lateral compressibility \cite{Heimburg1998, Ebel2001}. This is so because the fluctuations in enthalpy and the fluctuations in area are proportional. Furthermore, the lifetime of the fluctuations is also proportional to the excess heat capacity \cite{Grabitz2002, Seeger2007}.  In agreement with this, previous experimental results suggest that the open likelihood and the open lifetime of lipid channels are directly related to fluctuation amplitudes and lifetimes. It implies, that in melting transitions the likelihood of finding pores in the membranes is high, and the open lifetime is long \cite{Blicher2009, Wunderlich2009, Gallaher2010}. The latter effect is known as `critical slowing-down'. The channel-like conduction events in membranes close to transitions just reflect the thermal fluctuations in the membrane, and the lifetime of the lipid channels corresponds to the fluctuation or relaxation time-scale, respectively \cite{Seeger2007}. This is important because biological membranes under physiological conditions in fact exist in a state close to melting transitions \cite{Heimburg2005c, Heimburg2007a, Heimburg2007c}, and lipid-based channels can be expected under physiological conditions. 

Recently, we have shown that the channel events in artificial membranes are strikingly similar to protein conduction (\cite{Laub2012}, demonstrated in comparison to various TRP channels) such that they cannot be distinguished by just investigating the traces. This finding suggests the possibility that  the origin of ion channel events may not always be correctly assigned in the literature. The main difference between membranes in the absence and presence of channel proteins is that the membranes containing proteins coherently show a well-defined value for the single-channel conductance, while the channel conductance in pure lipid membranes is subject to considerable variations between individual experiments. The reason for this is presently not clear, but may originate from sample preparation and the influence of the pipette walls on the phase behavior of membranes. 

Membrane transitions can be influenced by changes in various thermodynamic variables, e.g., hydrostatic and lateral pressure, calcium concentration (chemical potential of Ca$^{2+}$), pH (chemical potential of H$^+$), but also by small drugs such as general or local anesthetics \cite{Heimburg2010}. Consequently, lipid channels are sensitive to anesthetics \cite{Blicher2009}. They can be blocked by anesthetics (such as octanol). This implies that drugs can act on channel events in the absence of any specific macromolecular receptor molecule. Rather, their effect is of a purely thermodynamic nature known as `freezing-point depression' \cite{Heimburg2007c}.

It is a common observation in the lab that lipid ion channels can be induced by voltage. Typically, the membrane ruptures at high voltage. At voltages somewhat below the rupture voltage one can induce channel formation \cite{Laub2012}. In order to systematically study this effect in a quantitative manner one needs synthetic membrane preparations that are stable over a significant period of time in a large range of voltages. Close to transitions, this happens only rarely. Using black lipid membranes, Wodzinska et al. \cite{Wodzinska2009} showed the voltage dependence of channel formation in a dioleyl phosphatidylcholine:dipalmitoyl phosphatidylcholine=2:1 mol:mol mixture in a range from -220 mV to +220V. More than 300 individual traces of 30 second length could be analyzed. In this experiment, a symmetric current-voltage profile and a linear single-channel conductance of $\gamma \sim$ 330 pS was found. The current traces clearly indicated that the channel open likelihood is a function of voltage (Fig. 2 in \cite{Wodzinska2009}). While only very few open events were found at voltages below 40mV, the frequency and duration of channel open events increased significantly at higher voltages until the channels are open most of the time. While this represents a proof-of-concept for the voltage-gating of lipid pores, the authors unfortunately did not analyze the voltage dependence of the open probability distribution in more detail.

In this publication, we demonstrate quantitatively that lipid channels may be gated by voltage. The effect of voltage on black lipid membranes was investigated by several authors in the 1970s. It was found that charging of the membrane capacitor generates a force on the membrane that leads to an effective reduction of the membrane thickness and thereby to an increase in capacitance with a quadratic voltage dependence. It has been proposed that electroporation (the generation of voltage-induce pores in bilayers) is a consequence of these forces \cite{Crowley1973, Abidor1979a, Winterhalter1987, Neumann1999}. Recently, we proposed that the electrostatic forces associated with an excess charge can also induce melting transitions in membranes \cite{Heimburg2012}.  Such transitions would influence the elastic properties of the membrane and thereby the likelihood of pore formation in the membrane \cite{Blicher2009}. Further, due to the coupling to fluctuation lifetimes, voltage can influence the open timescales of the lipid channels. In previous reports we have demonstrated that membranes can display non-linear current-voltage relationships \cite{Wodzinska2009, Laub2012} and that lipid channels can be induced by voltage. Here, we study the voltage-gating of lipid channels more systematically and propose a theoretical description for inward or outward rectified I-V curves of synthetic lipid membranes that is in excellent agreement with experimental data. 


\section*{Materials and Methods}
Lipids (1,2-dimyristoyl-sn-glycero-3-phosphocholine - DMPC; 1,2-dilaureoyl-sn-glycero-3-\linebreak phosphocholine - DLPC) were purchased from Avanti Polar Lipids (Alabaster, Al). In all experiments shown here, we used a DMPC:DLPC=10:1 mol/mol mixture. 
The 150 mM NaCl electrolyte solutions used for the patch clamp experiments were buffered with 2mM HEPES (Sigma-Aldrich, Germany) and 1mM disodium EDTA (Fluka, Switzerland). The pH was adjusted to 7.4 using NaOH. The specific conductivity of the solution was measured to be 1.412 S/m at 21$^\circ$C. In order to get rid of dust particles and other possible contaminants, the buffer was always pressed through a sterile 0.2$\mu$m filter (Minisart\textsuperscript{\textregistered} \#16532, Sartorius Stedim Biotech, France) immediately before use. All chemicals used were of the highest purity grade available and any water used was purified (resistivity $>$ 18M$\Omega\cdot$ cm) on a desktop EASYpure RF water purification system from Barnstead/Thermolyne (Dubuque/IA, USA). 

Each lipid type was dissolved separately in an organic solvent (dichloromethane : metha\-nol=2:1). The two components were then mixed in the desired ratio. This mixture was dried by gentle heating while being exposed to an air stream. The remaining organic solvent was then removed by placing the sample in a high vacuum desiccator for a couple of hours. The resulting lipid film was then resuspended by adding a highly volatile solvent (hexane:ethanol 4:1 by volume). The final lipid concentration in the solvent was $\sim$ 1-2mM. 

Electrophysiological measurements were performed on single synthetic membranes span\-ned over the tip of a patch pipette using the tip-dipping method \cite{Hanke1984}. In this method, the planar lipid bilayer membrane is formed on the tip of a patch-clamp glass pipette that has been filled with electrolyte solution. The tip is in contact with the surface of a beaker filled with the same electrolyte solution. The lipid solution in the hexane/ethanol mixture is then brought into contact with the outer surface of the glass pipette. When the solution flows down the pipette, a membrane spontaneously forms at the tip of the pipette. The solvent was allowed to evaporate for at least 30 seconds before the pipette was lowered 4Ð5mm below the bath surface. The main advantage of this method is that the resulting membrane is practically solvent free.

The patch diameter was $\sim$\,1 $\mu$m. Pipettes were pulled from 1.5 mm/0.84 mm (outer diameter/inner diameter) borosilicate glass capillaries (\#1B150F-3, World Precision Instruments, USA) with a vertical PC-10 puller from Narishige Group, Japan. Current recordings were made using an Axopatch 200B patch clamp amplifier (Axon Instruments Inc., Union City/CA, USA). The pipette and electrodes were mounted on a cooled capacitor feed- back integrating headstage amplifier (Headstage CV 203BU, Axon Instruments Inc.). The same experimental procedure was used and described in \cite{Laub2012}.

Channel kinetics has been determined by defining a suitable threshold for an open event in an automated procedure. The threshold was set to be $\pm 4\sigma$, where $\sigma$ is the noise of the baseline current.  The method is described in more detail in the PhD thesis of one of the authors \cite{Blicher2011}. Somewhat more sophisticated methods to bin the kinetic data in a logarithmic manner can be found in \cite{Sigworth1987, Stark2000, Liebovitch2001}.

The patch-pipette experiments have the advantage over black lipid membrane experiments that they result in much more stable and consistent conductance traces, and the membranes don't rupture so easily. However, due to the small patch size, there seems to be a large influence of the pipette on the membranes. Newly formed membranes over fresh pipettes lead to somewhat varying quantitative results. The voltage dependence of our membranes was therefore always recorded for one unique membrane and one pipette. While trends and qualitative behavior are reproducible between different experiments, the absolute numbers vary (see \cite{Laub2012} for more details). For instance, the currents shown in Figs. \ref{Figure2} and \ref{Figure3c} display very similar qualitative behavior, while the absolute conductances and the associated free energy changes are somewhat different. Our experiments provide a qualitative proof-of-principle.
\begin{figure*}[t!]
    \begin{center}
	\includegraphics[width=16.0cm]{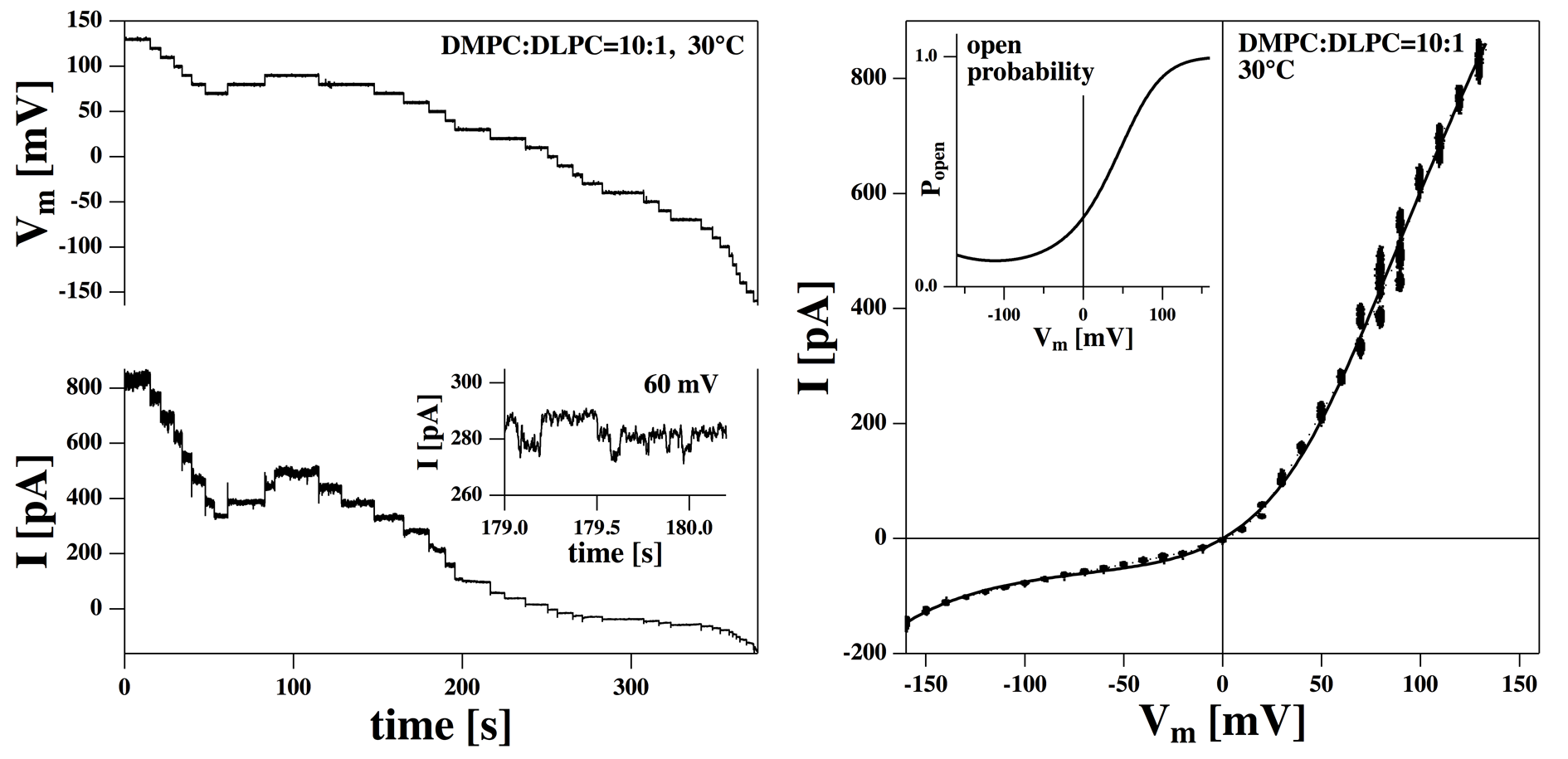}\\
	\parbox[c]{15cm}{\caption{\small\textit{Current-voltage profile of a DMPC:DLPC=10:1 mol:mol membrane at 30$^\circ$C.} Left: Membrane voltage and the corresponding total membrane current as a function of time. The voltage was changed stepwise during the experiment. The insert shows some small channel events at V$_m$=60 mV with a conductance of about $\gamma$ = 125 pS. Right: Resulting current-voltage relation. The I-V curve is outward rectified. The solid line is a fit to eq. \ref{eq:05} (parameters are given in the text). The insert shows the corresponding open probability of a membrane pore deduced from eq. \ref{eq:04}. It displays a minimum at $V_0 = -110 mV$.}
	\label{Figure1b}}
    \end{center}
\end{figure*}
\section*{Theory}

The electrostatic force, $\mathcal{F}$, exerted on a planar membrane by a voltage is given by 
\begin{equation}
\label{eq:01}
\mathcal{F}=\frac{1}{2}\frac{C_m V_m^2}{D}
\end{equation}
where $C_m$ is the membrane capacitance, $V_m$ is the transmembrane voltage and $D$ is the membrane thickness \cite{Heimburg2012}. This force reduces the thickness of the membrane \cite{White1973}. The electrical work performed on the membrane by a change in thickness from $D_1$ to $D_2$ is
\begin{equation}
\label{eq:02}
\Delta W_{el}=\int_{D_1}^{D_2}\mathcal{F}dD=\frac{1}{2}\epsilon_0 V_m^2\int_{D_1}^{D_2}\epsilon \frac{A}{D^2}dD\equiv \alpha V_m^2
\end{equation}
where $A$ is area, $\epsilon_0$ is the vacuum permittivity, and $\epsilon$ is the dielectric constant of the membrane core. The coefficient $\alpha$ is constant for constant temperature and pressure. This work is proportional to the square of the voltage, and one would thus reasonably assume that the free energy of pore formation is related to the square of the voltage and to the elastic constants of the membrane. This has been proposed previously \cite{Winterhalter1987, Glaser1988}. Therefore, the free energy for pore formation has the form
\begin{equation}
\label{eq:03}
\Delta G=\Delta G_0 +\alpha V_m^2
\end{equation}
where $\Delta G_0$ is the free energy difference between open and closed pore in the absence of voltage. $\Delta G_0$ reflects the elastic properties of the membrane that depend on composition, temperature and pressure. Such quadratic dependence is reasonable for symmetric membranes because a linear term can be excluded for symmetry reasons. For asymmetric membranes one obtains
\begin{equation}
\label{eq:03b}
\Delta G=\Delta G_0 +\alpha (V_m-V_0)^2 \;,
\end{equation}
where the offset voltage $V_0$ may be due to membrane curvature, the presence of the pipette wall on one side of the membrane or to a different lipid composition in the two membrane leaflets\cite{Alvarez1978}. Membrane curvature could possibly originate from slight pressure difference on the two sides of the membrane due to suction or dipping the pipette tip deeper into the aqueous medium (see Materials section). At the end of the 'Results and Discussion' section we outline how such a curvature can create a transmembrane voltage on the order of 100 mV associated to a phenomenon called flexoelectricity.

The probability, $P_{open} (V_m)$, of finding an open pore in the membrane at a fixed voltage is given by
\begin{equation}
\label{eq:04}
P_{open}(V_m)=\frac{K(V_m)}{1+K(V_m)}\;;\;K(V_m)=\exp\left(-\frac{\Delta G}{kT}\right) \;,
\end{equation}
where $K(V_m)$ is the voltage-dependent equilibrium constant between open and closed states of a single pore. 

The current-voltage relation for the lipid membrane is proportional to the likelihood of finding an open channel for a given voltage:  
\begin{equation}
\label{eq:05}
I_m=\gamma_p\cdot P_{open}\cdot (V_m-E_0)\;,
\end{equation}
where $\gamma_p$ is the conductance of a single pore (or N identical pores) and $E_0=$ is the Nernst potential of a particular ion with charge $z$ given by $E_0=(RT/zF)\ln ([C_{out}]/[C_{in}])$.
While the voltage $V_0$ reflects the asymmetry of the membrane, $E_0$ reflects the asymmetry of the ion concentrations of the buffer solution. If the aqueous buffer is the same on both sides of the membrane, the Nernst potential is zero.  Eqs. \ref{eq:03b}-\ref{eq:05} contain the theoretical description for the I-V curves of lipid channels. It is based on the concept of forces induced by charging the membrane capacitor.\\

The above simple capacitor model differs from the commonly used Eyring-type activation scheme of form
\begin{equation}
\label{eq:Eyring1}
\resizebox{.912\hsize}{!}{$I=z F k_0 \left[[C]_{in}\exp\left(\frac{\delta z F V_m}{RT}\right) - [C]_{out}\exp\left(-\frac{(1-\delta) z F V_m}{RT}\right)  \right]$}
\end{equation}
that assumes flow of charges in a field and a linear dependence of the free energy of the conduction rates on voltage. Here, $\delta$ is the position of an activation barrier in the membrane. The Hodgkin-Huxley gate model makes use of such an activation scheme \cite{Johnston1995}. 

\section*{RESULTS AND DISCUSSION}

We performed patch recordings on a DMPC : DLPC = 10:1 membrane \cite{mixture} at T=30$^{\circ}$C in a 150mM KCl buffer. This mixture displays a chain melting transition close to room temperature \cite{Laub2012}. 
Fig. \ref{Figure1b} (left) shows a time sequence of the changes in total membrane current when voltage is changed stepwise. Around 50-60 mV one finds small quantized steps on the current trace which are not always well resolved (insert to Fig. \ref{Figure1b}, left). The single channel conductance of these steps is $\approx$ 125 pS. This value is small compared to the overall conductance of the total membrane at this voltage of $\approx$ 4.7 nS. At other voltages, no single channels were observed. The asymmetry of the membrane becomes apparent in Fig. \ref{Figure1b} (right), which shows the current-voltage relation for the total membrane current (i.e., all current including the baseline current) of the membrane in Fig. \ref{Figure1}. The solid line represents a fit to eq. \ref{eq:05} yielding $V_0=-110$mV, $\gamma_p=6.62$ nS, $\Delta G_0=5.2$ kJ/mol, and $\alpha=-248$ kJ/mol$ \cdot$V$^2$). This fit reproduces the experimental current-voltage profile very well. Such an outward rectified I-V curve for a synthetic membrane was already shown in \cite{Laub2012} where it was analyzed with an Eyring transition state model (eq. \ref{eq:Eyring1}) that is very commonly used in the protein channel field. The present theoretical description yields a much superior description of the experimental data.  Eq. \ref{eq:05} is based on the assumption of a single channel that opens in a voltage-dependent manner. The open likelihood of this channel (eq. \ref{eq:04}) is shown in the insert to the right hand panel of Fig. \ref{Figure1b}.  The present theoretical treatment also yields very good  descriptions of the I-V profiles of various transient receptor potential (TRP) channels reconstituted in human embryonic kidney cells with very similar parameters as used for the synthetic membrane (cf., \cite{Laub2012}, fits shown in \cite{Mosgaard2013a}). The I-V curves of the TRP channels were also described with much less success by the Eyring model \cite{Laub2012}. 

\begin{figure}[htb!]
    \begin{center}
	\includegraphics[width=8.5cm]{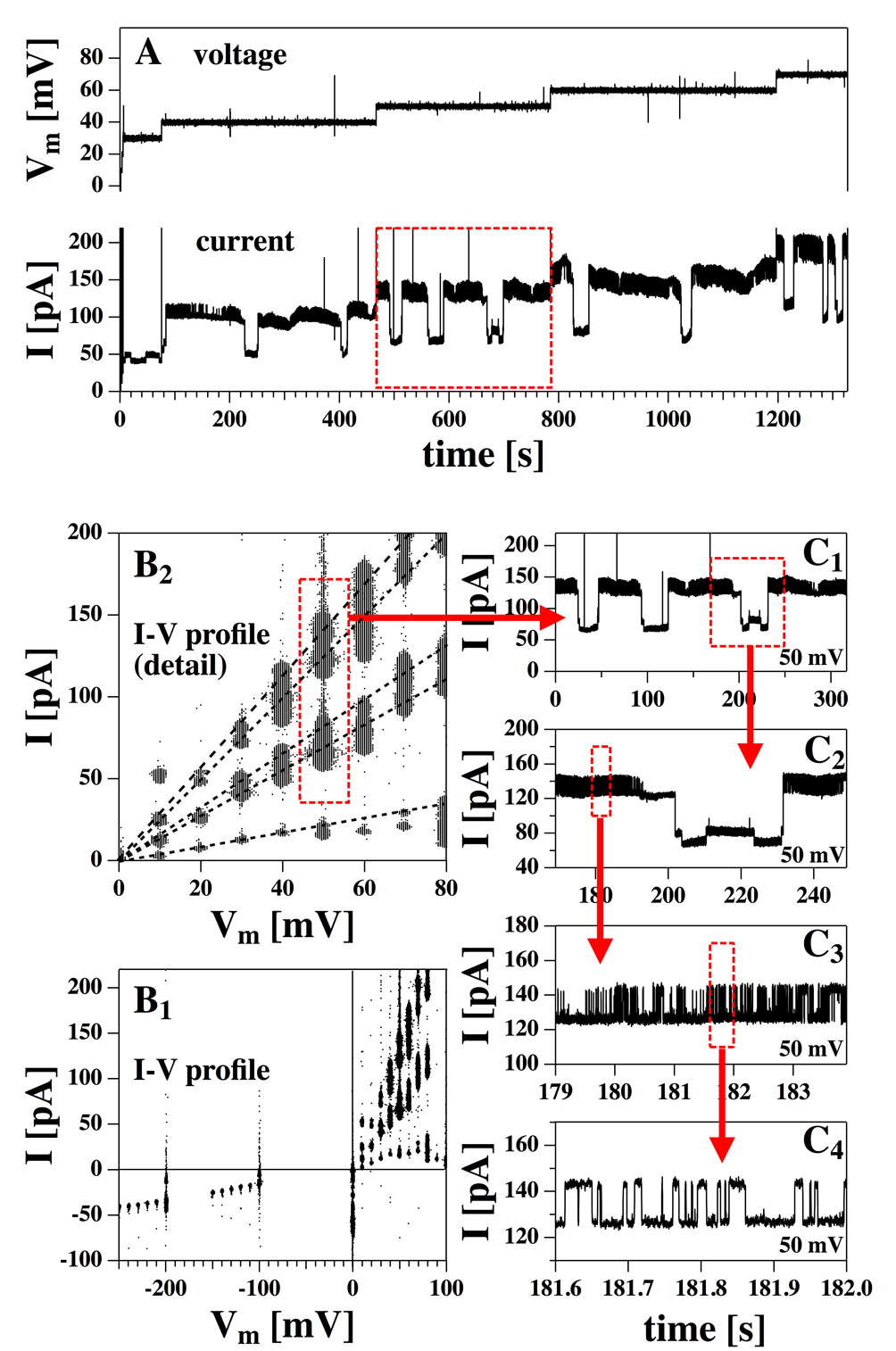}
	\parbox[c]{8cm}{\caption{\small\textit{Current-voltage behavior of a different preparation of a DMPC:DLPC=10:1 mol:mol membrane at 30$^\circ$C showing many channel events.  A: Section of the raw voltage and current traces (20 minutes out of 1 hour recording). The section in the red box is amplified in panel C$_1$. B$_1$: I-V profile of the complete data set. At positive voltages one can recognize $\sim$5 individual current levels. B$_2$: Amplified section of panel B$_1$. The black dashed lines are guides to the eye to mark the five current levels. The red box marks the four current levels that can be seen in panel C$_1$. C$_1$: Current I at V$_m$=50mV over 5 minutes. Panels C$_2$-C$_4$: Consecutive enlargement of small sections of the profile in panel C$_1$. In panel C$_4$, two individual currents levels can easily be recognized. These two levels are investigated in more detail in Fig. \ref{Figure1}.}}
	\label{Figure1c}}
    \end{center}
\end{figure}
Two conclusions can be drawn from the I-V curve (Fig. \ref{Figure1b}): 
\begin{enumerate}
  \item There are net currents through the membrane even under conditions where individual channels cannot be visually resolved. These currents are well-described by a simple model that assumes the formation of pores with well-defined size. If such pores in fact exist, they must be short lived or much smaller than the relatively big steps in Figs. \ref{Figure1c}, \ref{Figure1} and \ref{Figure3c} (discussed below). 
  \item  In this particular experiment, the offset-voltage of the membrane in the patch pipette is $V_0=-110 mV$, which indicates that the membrane is asymmetric. Below, we will use this offset voltage to describe the single channels in Fig. \ref{Figure1}. In the BLM experiment by Wodzinska et al. \cite{Wodzinska2009} discussed in the introduction, $V_0$ was $\approx 0$ mV, indicating a symmetric membrane. It is a general observation in our lab that BLM measurements on large membranes tend to result in more symmetric current-voltage relations than patch recordings with small tips diameters.
\end{enumerate}
\begin{figure}[t!]
    \begin{center}
	\includegraphics[width=8.5cm]{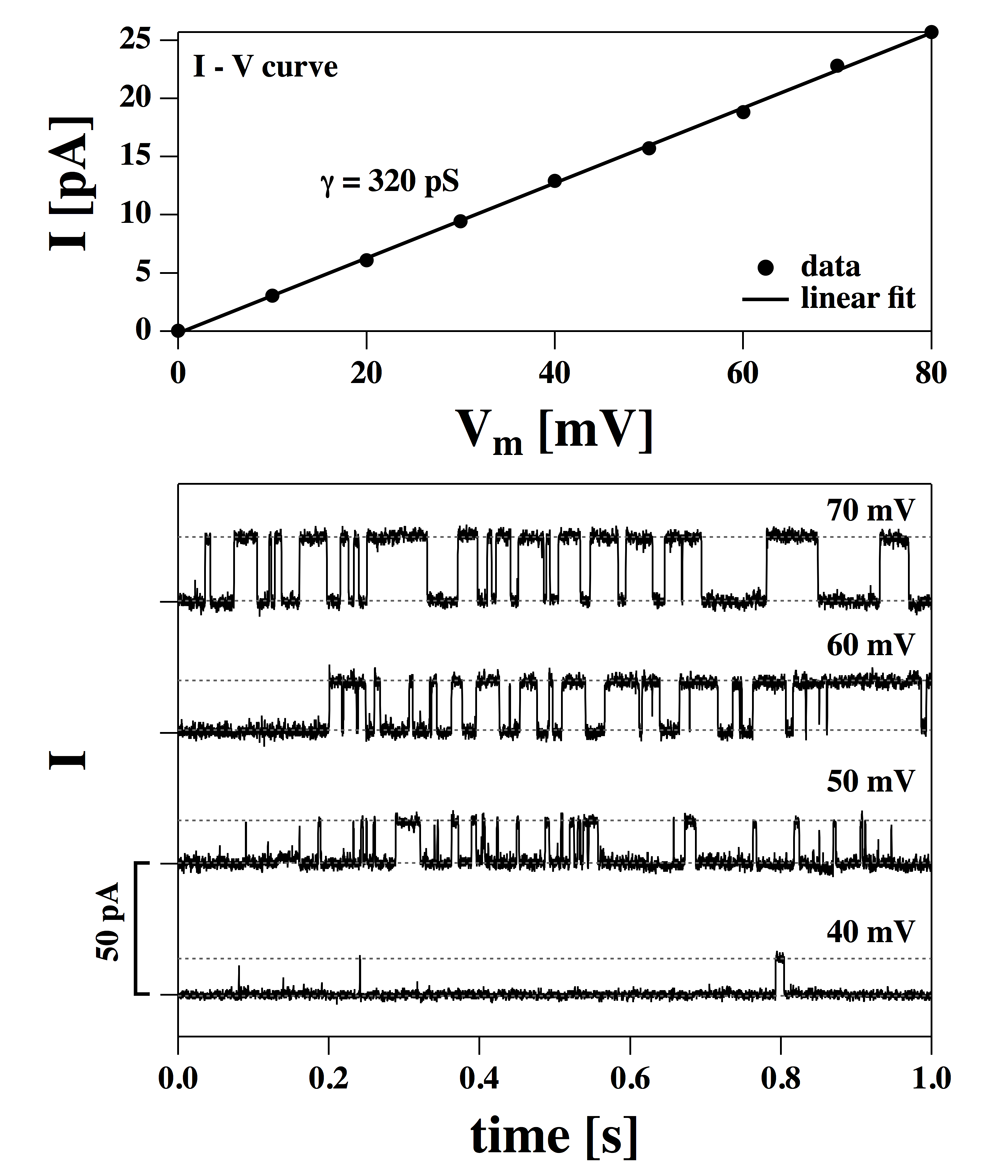}
	\parbox[c]{8cm}{\caption{\small\textit{Recordings of the DMPC:DLPC=10:1 membrane from Fig. \ref{Figure1c} as a function of voltage (cf. Fig. \ref{Figure1c}, panel C4.)}  Bottom: Current-traces for a DMPC:DLPC=10:1 mol/mol membrane (150mM KCl, T=30$^{\circ}$C) at four voltages showing an increase of single-channel conductance with voltage and an increased likelihood of channel formation. Top: The corresponding linear single-channel current-voltage relation indicating a single-channel conductance of $\gamma$=320 pS.}
	\label{Figure1}}
    \end{center}
\end{figure}

Fig. \ref{Figure1b} also demonstrates an experimental problem with obtaining statistically relevant data: The membrane has to be stable over the whole voltage regime of interest, and the overall properties have to remain constant during the total time of the experiment. In Fig. \ref{Figure1b}, the currents at voltages between +70 mV and +90 mV were recorded three times within two minutes. The variations during this time show up in the large noise in the respective interval in the I-V curve. Obviously, the membrane properties drift slightly during the experiment. While the voltage-gating of the lipid channels is a generic phenomenon whenever lipid channels can be recorded, it is rather rare that a complete I-V curve can be investigated over a long period without drifts or breaking of the membrane. 

Fig. \ref{Figure1c} shows another experiment performed at the same conditions as the one in Fig. \ref{Figure1b}. Panel A contains a section of about 25 minutes of the time dependence of both voltage and membrane current (the total experiment lasted more than an hour). Panels B$_1$ and B$_2$ (containing a magnification of B$_1$) show the I-V plot using the total raw data (V$_m$ ranging from -250 mV to +100 mV). At positive voltages, one can recognize individual peaks in the distribution, corresponding to at least five individual current steps. Four events at V$_m$ = 50 mV are highlighted in the red box in panels A and B$_2$. The corresponding amplified current trace is shown in panel C$_1$ and consecutive enlargements are shown in panels C$_2$ to C$_4$.  One can clearly recognize four individual current levels, which must be attributed to at least 2-3 different pores with open and closed states. The difference in current between these four steps corresponds to three conductances of $\gamma \sim$ 240pS, $\gamma \sim$ 900 pS and $\gamma \sim$ 320 pS (extracted from panel C2). 
\begin{figure}[t!]
    \begin{center}
	\includegraphics[width=8.5cm]{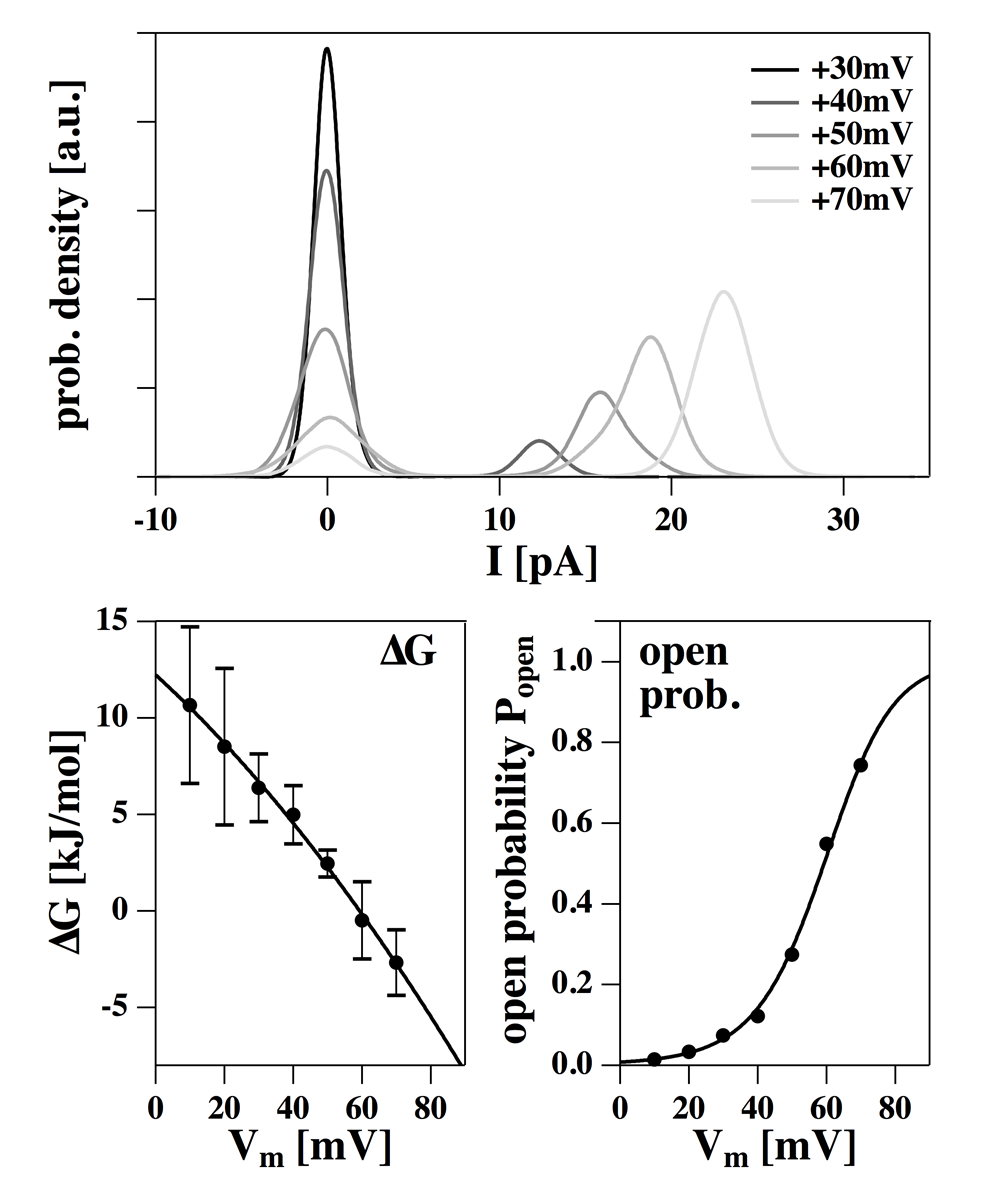}
	\parbox[c]{8cm}{\caption{\small\textit{Statistical analysis of the data in Fig.~\ref{Figure2}.}  Top: Experimental probability distribution for channel opening at five voltages (cf. Fig. \ref{Figure1}).  Bottom, left: The calculated free energy difference between open and closed state as a function of voltage indicates a quadratic voltage dependence (see eq. \ref{eq:03b}, $\Delta G_0$=21.3 kJ/mol; $\alpha$=-721 kJ/ mol$\cdot$V$^2$, V$_0$=-110 mV). Bottom, right: The open probability, $P_{open}$, shows a transition at $\sim$60 mV.}
	\label{Figure2}}
    \end{center}
\end{figure}

The data in Fig. \ref{Figure1} (bottom) represent the two current steps with the largest conductance shown in Fig. \ref{Figure1c} (cf, panel C$_4$), recorded at four different voltages. The recordings were quite stable showing permanent opening and closing of channels over up to 30 minutes for a single voltage. Increasing the voltage leads to a larger single-channel current and to an increased likelihood of finding open channels. The current-voltage relation for the single-channel current is linear (Fig. \ref{Figure1}, top), indicating the presence of lipid channels with a constant pore size. The single-channel conductance is  $\gamma=$ 320 pS and does not seem to be voltage-dependent. We could not resolve any single-channel activity at negative voltages indicating an asymmetric membrane. The traces allow for an accurate statistical analysis of the single channel events. Fig. \ref{Figure2} (top) shows the current histograms associated with the current traces of Fig. \ref{Figure1}. One finds two peaks in the distribution representing the closed and the open states. The current of the closed state was set to zero. The (voltage-dependent) equilibrium constant, $K(V_m)$, between open and closed state can be deduced from the peak areas of the two states in the histogram. From the equilibrium constant one can deduce the free energy difference of the two states (Fig. \ref{Figure2}, bottom left) and the open probability  (Fig. \ref{Figure2} bottom right) --- see eq.\ref{eq:04}. The solid lines in the bottom panels represent fits assuming a quadratic form of the free energy as given by eq. \ref{eq:03b} with an offset voltage of $V_0=-110 \pm 60$ mV, $\Delta G_0$=21.3 kJ/mol and $\alpha$=-721 kJ/mol$\cdot$ V$^2$. V$_0$ is consistent with the offset voltage obtained from the I-V profile in Fig. \ref{Figure1b}. It is also in agreement with the observation that one does not find any channel events at negative voltages in the regime investigated here. $\Delta G_0$ and $\alpha$ assume larger values as for the I-V profile in Fig. \ref{Figure1b} which indicates that the pores are larger. Fig. \ref{Figure2} indicates that one finds a voltage-induced transition from a closed to an open state over a range of about 30 mV. Such a range is also typical for the activation protein channels. For instance, the voltage-gated K$^+$-channel HERG \cite{Trudeau1995} displays an closed-open transition with a similar width \cite{Trudeau1995}.  If one attributes the open-closed event to a single protein, one typically would associate it to a gating charge moving in a field. The open-probability shown in Fig. \ref{Figure2} (bottom right) can be reasonably well described by such a model. The equilibrium between a closed and an open channel described by  a gating charge $z$ is given by $K=\exp[-(\Delta G_0+z F V_m)/RT]$. The best fit yields  $z\approx 2.5$, which is of the same order as values found for many proteins. However, as we have shown in \cite{Heimburg2012} and in the theory section, the gating could also originate from charging the membrane capacitor as a whole. 
\begin{figure*}[ht!]
    \begin{center}
	\includegraphics[width=16cm]{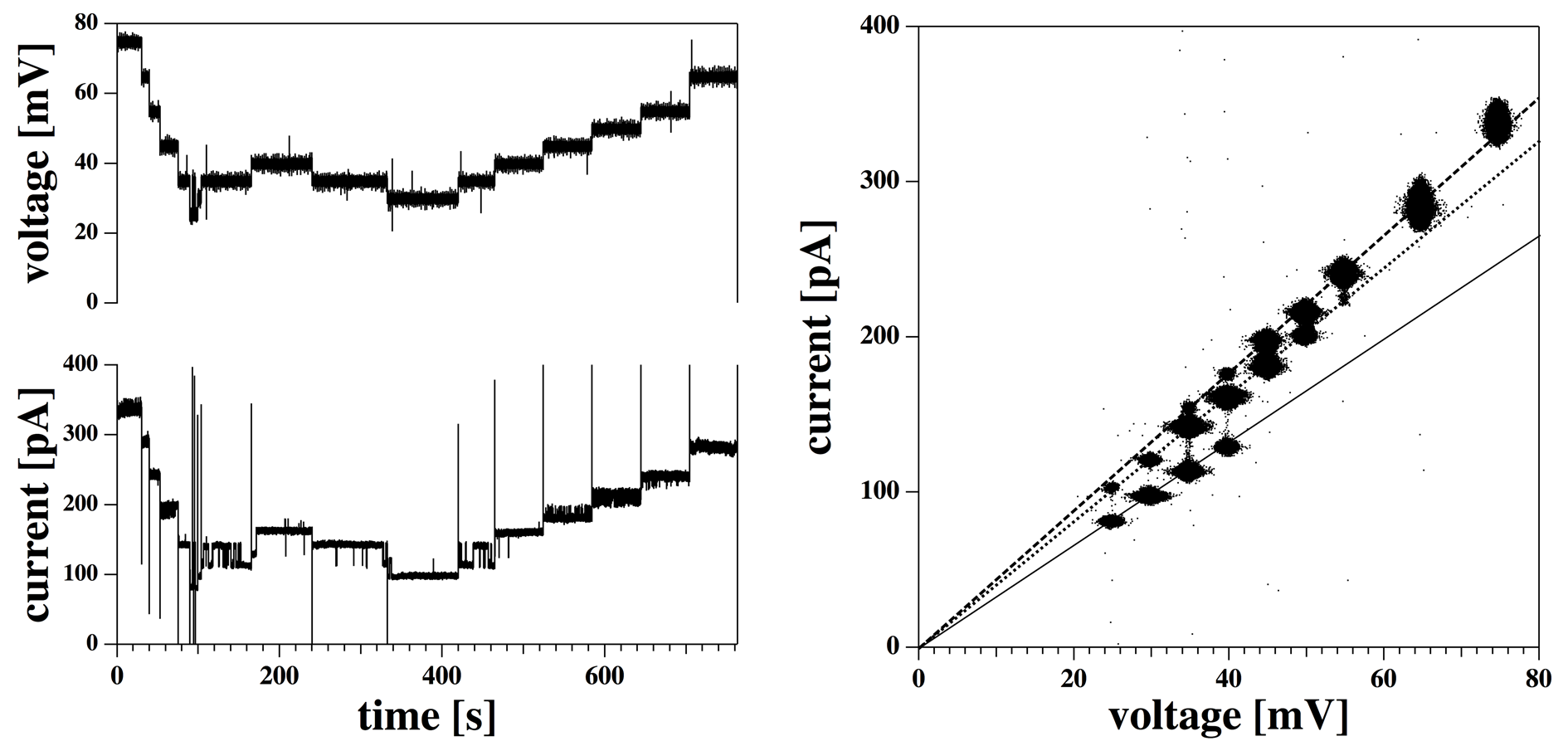}
	\parbox[c]{15cm}{\caption{\small\textit{Recordings on a DMPC:DLPC membrane at 29$^\circ$C.}  Top: Probability distribution for the first (left) and the second open state (right). Bottom: The calculated free energy difference for both conduction steps is consistent with similar quadratic voltage dependence (for parameters see text). }
	\label{Figure3b}}
    \end{center}
\end{figure*}

\subsection*{Traces with two steps}
In Fig. \ref{Figure3b} the voltage and current traces of a different membrane  (DMPC:DLPC=10:1; T=29$^{\circ}$C) are displayed together with the corresponding I-V plot (right panel), recorded at positive voltages only. In this plot, three current levels can nicely be resolved. At some voltages (30 and 40 mV) these pores coexist. As in the previous examples,  the current at each of these levels displays a linear current-voltage relation (indicated by the straight dashed lines in Fig. \ref{Figure3b}, right). However, the likelihood to find any one of these levels is voltage-dependent.

\begin{figure*}[t!]
    \begin{center}
	\includegraphics[width=16.0cm]{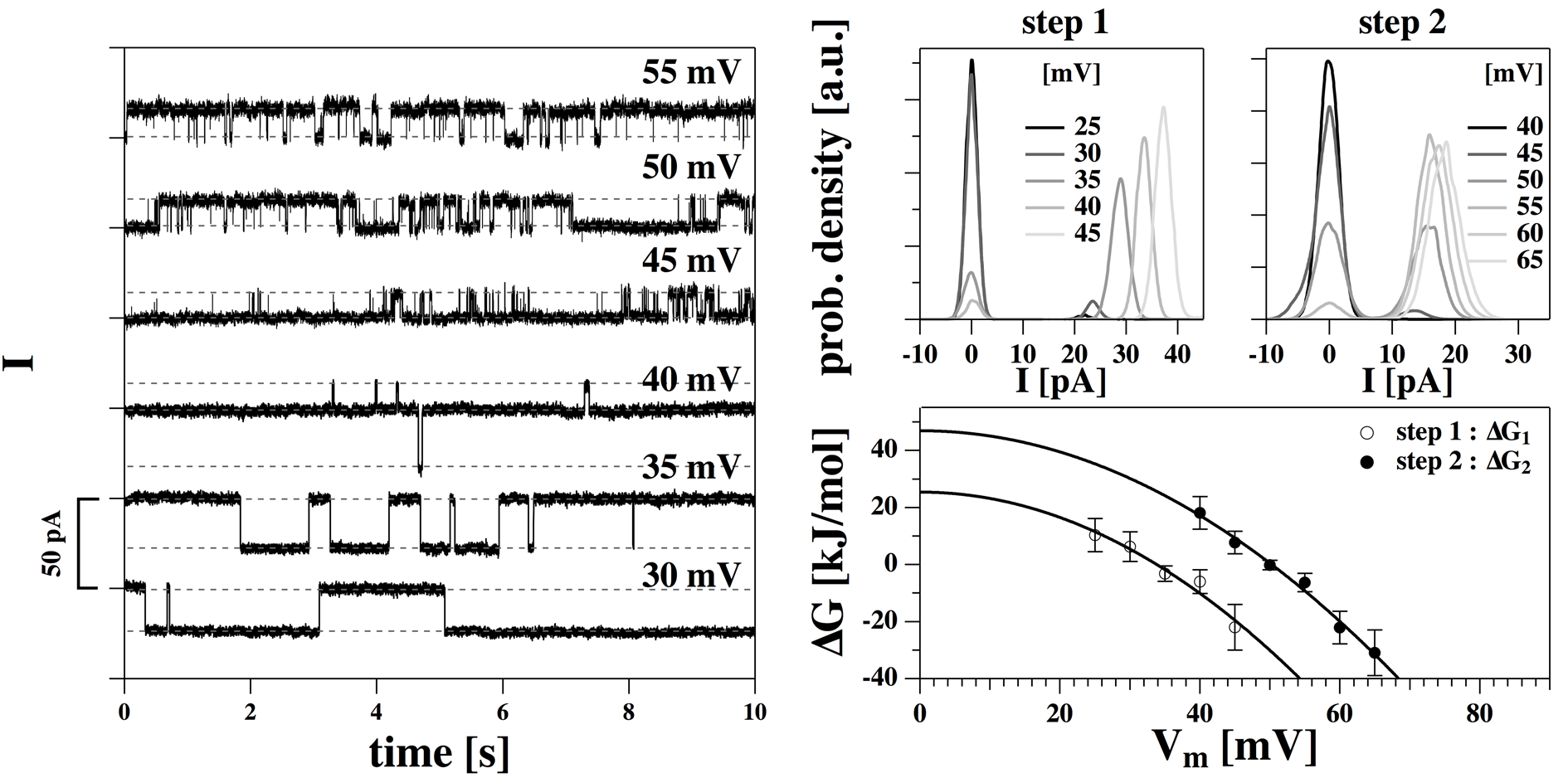}\\
	\parbox[c]{15cm}{\caption{\small\textit{Recordings on a DMPC:DLPC membrane at 29$^\circ$C.}  Left: Current traces at 6 voltages from 30mV to 55mV. THe trace at 40mV shows twos large and a small current step. Right, top: Probability distribution for the first (left) step with a conductance of $\gamma$=830 pS and the second open state (right) with a conductance of $\gamma$= 287 pS. Bottom: The calculated free energy difference for both conduction steps is consistent with a nearly identical quadratic voltage dependence ($\Delta G_{0,1}$=25.4 kJ/mol; $\alpha_1$=-22.2$\cdot $10$^{3}$ kJ/ mol$\cdot$V$^2$; $\Delta G_{0,2}$=46.9 kJ/mol; $\alpha_2$=-18.6$\cdot $10$^{3}$ kJ/ mol$\cdot$V$^2$).}
	\label{Figure3c}}
    \end{center}
\end{figure*}

Fig. \ref{Figure3c} (left) shows details of the current trace in Fig. \ref{Figure3b} at 6 different voltages from 30 to 55 mV. In the 40 mV trace one can recognize two steps of different magnitude. At lower voltages, a large pore with a single-channel conductance of $\gamma \sim$ 831 pS is found, whereas at high voltages one finds a different pore with $\gamma \sim$ 287 pS. Overall, the high voltage channel has a smaller single-channel conductance as the low voltage channel. As in Fig. \ref{Figure1c} the large pore displays longer open and closed times. The right hand panel shows the histograms of the first and the second voltage step as a function of voltage. Zero pA corresponds to the baseline current. The closed-to-open transition of the first step takes place at $\sim$34 mV. The transition of the second step is found at $\sim$50 mV. Here, zero pA corresponds to the current of the first open channel. The free energy of the two steps  as a function of voltage is given in Fig. \ref{Figure3c} (bottom right). The free energy of step 2 represents that of the single channel (i.e., it is the difference in free energy between the first and second open event). Both curves are well approximated by quadratic functions with V$_0$=0 mV. At zero voltage, the free energy of the second pore ($\Delta G_{0,2}$) has been found to be approximately twice the free energy of one pore ($\Delta G_{0,1}$=25.4 kJ/mol; $\alpha_1$=-22.2$\cdot $10$^{3}$ kJ/ mol$\cdot$V$^2$; $\Delta G_{0,2}$=46.9 kJ/mol; $\alpha_2$=-18.6$\cdot $10$^{3}$ kJ/ mol$\cdot$V$^2$). This is also reflected in the fact that the second pore forms at a higher voltage. The parameter $\alpha$ describing the voltage dependence of each pore is the same within error.  

As mentioned already in the Materials section, our patch pipette method yields qualitative similar results in recordings for different membranes or methods. We found similar channel events in BLM recordings, when using patch pipettes with sharp edges, and with fire-polished pipettes. This indicates that the appearance of the channels as such is not primarily due to the nature of a particular interface. However, it is possible that channel probabilities are altered by the presence of pipette or Teflon interfaces or by other membrane perturbations, and we argued earlier that the domain interfaces and proteins could also represent such perturbations \cite{Heimburg2010}. All of these perturbations may locally alter membrane fluctuations. The experiments display significant quantitative variations between preparations. This implies, that lipid channels may display somewhat different conductances and voltage-dependences from experiment to experiment. The reason for this is not exactly clear. We have speculated that under the conditions used by us the lipid mixtures may display domains that are of similar size than the patch diameter. Domain sizes that are visible under a microscope are quite common in binary lipid mixtures close to melting transitions. This may lead to variations of the fluid membrane fraction in different experiment. As discussed above, another possibility is that the patch pipette itself has an influence on the physical behavior of the membranes as is known for the interaction of membranes with glass surfaces \cite{Yang2000}. A third possibility is that the membranes in the patch pipette are curved due to differences of pressure on both sides of the membrane. 
Petrov and collaborators have argued that membrane curvature is coupled to the generation of a transmembrane voltage $V_0$ due to a phenomenon called `flexoelectricity' \cite{Petrov2001}.
In the case of a spherically curved bilayer with radius of curvature $c$, Petrov finds that
 \begin{equation}
\label{eq:Petrov}
V_0 = \frac{2}{\epsilon_0}f\;c \;,
\end{equation}
 where $\epsilon_0=8.854\cdot 10^{-12} \, F/m$ is the vacuum permittivity and $f$ is a parameter determined from experiment \cite{Petrov2001}.
The value found by Petrov for egg yolk phosphatidylcholine in 0.1 \, M  KCl was $f=13\cdot 10^{-19}$ C (Coulombs). Consequently, an offset potential of the order of $V_0=100$ mV (as found in our present experiments) would be generated by a small radius of curvature of about $R=1/c=3 \mu m$. Such a curvature seems plausible for a membrane patch with a diameter of $\sim$1 $\mu$m in the presence of slight suction. A BLM with a diameter of 80 $\mu$m possesses a minimum radius of $R=40 \mu$m, which corresponds to a maximum voltage caused by flexoelectricity of $V_0=7$mV.  It is therefore expected that the offset voltage created by flexoelectricity is small for BLMs but putatively large for small patch pipettes as found by \cite{Wodzinska2009}. If pipette suction is not controlled, the channel open-probabilities would also not be well-defined, which may well explain the variations in our experiments.

It should be added the all of the above sources for the facilitation of ion channels are also present in patch-clamp experiments on biological preparations. For the above reasons, however, absolute numbers and the differences in conductances and free energies of pore formation for different membranes (as in Figs. \ref{Figure2} and \ref{Figure3c}) should not be over-interpreted. Within one preparation, however, all results are quantitatively reproducible. They all display voltage gating with a quadratic voltage-dependence of the open probability while simultaneously the single-channels conductances remain constant.
\begin{figure*}[t!]
    \begin{center}
	\includegraphics[width=12.0cm]{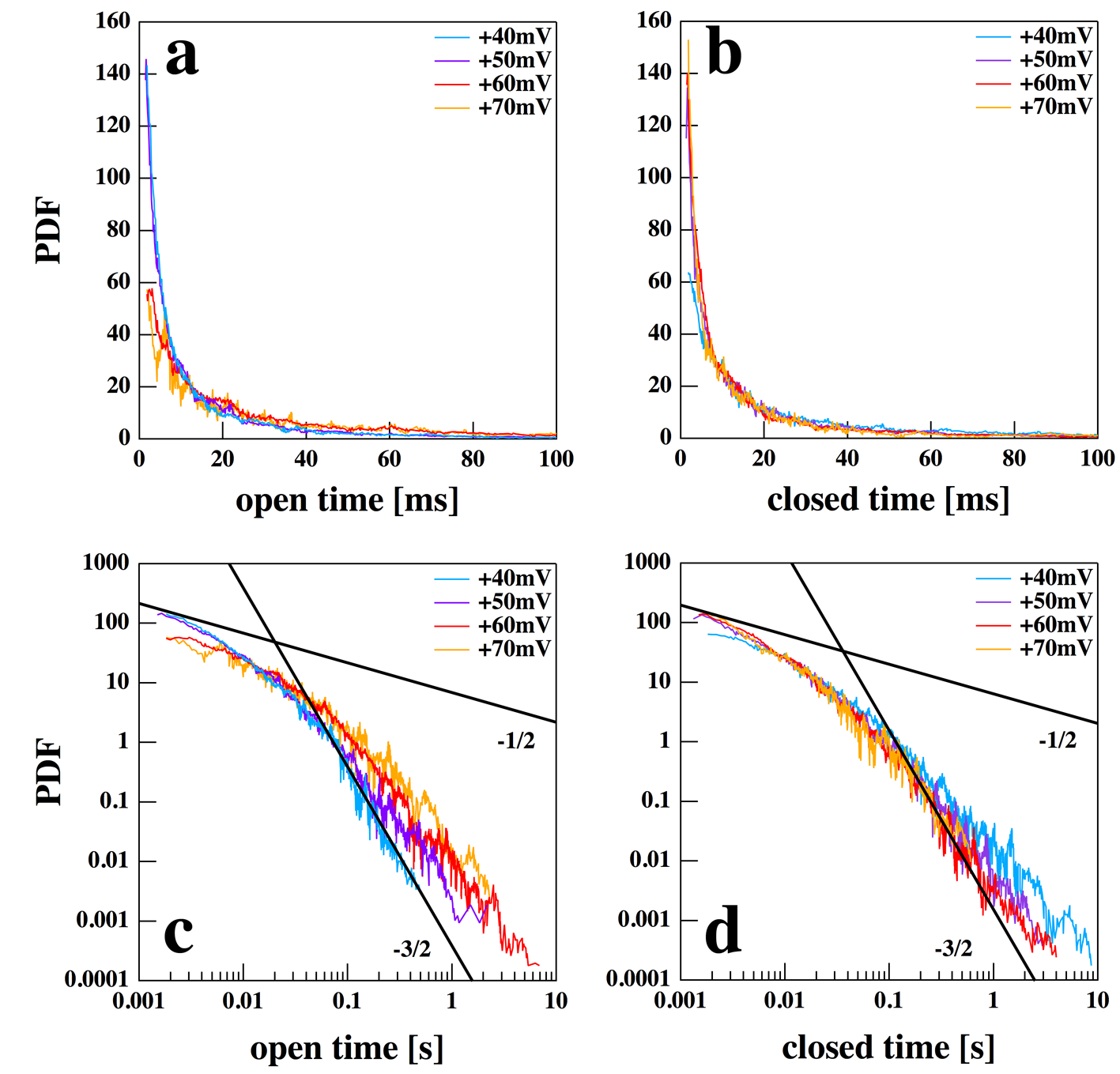}\\
	\parbox[c]{15cm}{\caption{\small\textit{Open and closed lifetimes of the lipid ion channels for the traces at four different voltages shown in Fig. \ref{Figure1}.} (A) Probability distribution functions (PDF) of the channel open-times at short times of up to 100 ms. (B) Closed time PDF. (C) and (D). Double-logarithmic representation of panels A and B over time scales up to 10 seconds. One can recognize that the PDFs are non-exponential, but also do not obey a simple power law in the double-log plot. The solid lines are guides to the eye with slopes of -1/2 and -3/2. At  long times the pdf seem to approach a $t^{-3/2}$ behavior. }
	\label{Figure4b}}
    \end{center}
\end{figure*}
\subsection*{Lifetime distributions}
The quite long channel traces shown in Figs. \ref{Figure1c} and \ref{Figure3b} allow for a detailed lifetime analysis of the lipid channels. In Fig. \ref{Figure4b} the probability distribution functions (PDF) of channel open- and closed-times is given for the channels shown in Fig. \ref{Figure1} (DMPC:DLPC=10:1; T=30$^{\circ}$C) at four different voltages. Panels \emph{A} and \emph{B} show a linear representation of the distributions for the open and closed lifetimes in the interval up to 100ms. In the interval up to 100ms these curves can seemingly be reasonably well described by a double-exponential function. For instance, a fit of the closed distribution at 60mV yields two lifetimes, 2.6 and 13.7 ms, respectively. However, closer inspection of the PDFs in a log-log representation (panels \emph{C} and \emph{D}) shows that the distributions range over several orders of magnitude and clearly display non-exponential behavior. Single-exponential behavior would be expected for statistical independent Markov-processes of two-state state conversions. However, at long open times, the slope in the open-time PDF approaches a  $t^{-3/2}$ power law behavior, which we found earlier in other synthetic lipid membrane preparations \cite{Gallaher2010}. A possible theoretical scheme for the open time PDF leading to such a power law given is 
\begin{equation}
\label{eq:07}
C\stackrel{k}{\longleftarrow} O_1\stackrel{K}{\rightleftharpoons} O_2 \stackrel{K}{\rightleftharpoons} O_3 \stackrel{K}{\rightleftharpoons} \quad...\quad \stackrel{K}{\rightleftharpoons} O_N
\end{equation}
where $C$ is a closed state and $O_1$, $O_2$, ... $O_N$ are N open states of a lipid pore with equal conductance, and the equilibrium constants between open states $K=O_{i+1}/O_i$ are constant. A similar fractal behavior was found also for proteins -  for instance for voltage-gated potassium channels \cite{Liebovitch1987c}. Originally, the above scheme was designed to describe such protein channels  \cite{Millhauser1988} by assuming $N$ open states of a channel protein. This means that  in respect to open-time distributions, lipid and protein channels are quite similar.  Our results show that fitting the open-time PDF with a bi-exponential function as commonly done for protein channels may be quite misleading. The PDFs show that such a description would largely underestimate the occurrence of very long open and closed events, respectively. 

A fractal nature of the PDFs must not necessarily originate from a discrete scheme as shown in eq. \ref{eq:07}. All of our experiments take place close to melting transitions, and it has be shown that the transition itself display fractal characteristics in terms of cooperative spacial density fluctuations \cite{Nielsen2000a} and most likely of temporal patterns, too.

The distribution profiles recorded at the four different voltages are seemingly very similar in the linear representation at short times. However, in the log-log representation it becomes obvious that especially the long open-times are significantly more frequent at higher voltage, while simultaneously the long closed events become less frequent. There is a finite probability to find very long open- or closed-times. The latter may give rise to silent periods in the conductance recordings followed by activity burst that have also been found in both in synthetic lipid membranes and protein channel-containing preparations \cite{Laub2012}.


\section*{CONCLUSIONS}
We have shown that lipid ion channels in model membranes are voltage-gated. This is in agreement with previous reports on the effect of electrostriction on membrane thinning \cite{Crowley1973, White1973, Heimburg2012}. The free energy of the open pore has a quadratic dependence on voltage. Such a quadratic voltage dependence is characteristic for the charging of a capacitor with electrical breakthroughs at critical voltages. It differs from a standard Eyring transition state model that considers movement of charges in a field. We could excellently describe the I-V profiles of membranes with our simple channel activation model when assuming an asymmetric membrane. The conductances of single channel events ($\gamma =$ 125 - 900 pS in the present experiments) and the closed-to-open transition width are of a similar order than the results found for protein channels. The size of the lipid pores does not seem to be fixed but varies between preparations. However, it is constant during one experiment. Our results complement earlier findings that lipid channels can be gated by other intensive variables, e.g., anesthetic drugs, temperature, pH, and calcium \cite{Heimburg2010}. The dependence of channel formation on these variables strongly suggests that these events are related to the well-understood physical chemistry of lipid membranes.  Increasing voltage favors channel openings with long time scales. 

The quantitative similarities between lipid and protein channels are intriguing. While they could be absolutely fortuitous, we find it more likely that these similarities are not accidental but rather suggest that conduction events have a common origin in the thermodynamics of the biomembrane.\\


\section*{Acknowledgments}
Acknowledgments: Andrew D. Jackson from the Niels Bohr International Academy kindly proof-read our manuscript. We thank one of the reviewers for pointing out the possibility of an asymmetric membrane. Karis Zecchi from the Membrane Biophysics Group provided the value for the offset voltage expected from flexoelectricity. A. B. was supported by a grant from the University of Copenhagen. T. H. was supported by the Villum foundation (VKR 022130).




%

\small{

}


\end{document}